\documentclass[12pt,a4paper]{amsart}

\usepackage[latin1]{inputenc}
\usepackage{amsmath,amssymb,amsthm,bm}

\theoremstyle{definition}

\newcommand{\hi}{\mathcal{H}} 
\newcommand{\ip}[2]{\left\langle\,#1\,|\,#2\,\right\rangle} 
\newcommand{\ket}[1]{|#1\rangle} 
\newcommand{\kb}[2]{|#1\rangle\langle#2|} 
\newcommand{\tr}{\textrm{tr\,}} 

\newcommand{\cal }{\mathcal}

\newcommand{\R}{\mathbb R}

\newcommand{\N}{\mathbb N}

\newcommand{\Ra}{\mathcal R}

\newcommand{\bor}[1]{\mathcal{B}({#1})}
\newcommand{\br}{\mathcal B(\mathbb R)}

\newcommand{\pom}{{\small\sf{POM }}} 
\newcommand{\Q}{\mathsf{Q}} 
\newcommand{\Eh}{\mathsf{E}_{\rm ht}} 
\newcommand{\W}{\mathcal{W}} 
\def\<{\langle}
\def\>{\rangle}

\newcommand{\sfq}{\mathsf{Q}}

\newcommand{\hB}{\mathcal{B}}

\begin{document}
\title[]{Continuous variable tomographic measurements}
\author{P. Lahti}
\address{Pekka Lahti,
Department of Physics and Astronomy, University of Turku,
FIN-20014 Turku, Finland}
\email{pekka.lahti@utu.fi}
\author{J.-P. Pellonp\"a\"a}
\address{Juha-Pekka Pellonp\"a\"a,
Department of Physics and Astronomy, University of Turku,
FIN-20014 Turku, Finland}
\email{juha-pekka.pellonpaa@utu.fi}
\begin{abstract}
Using a recent result of Albini {\em et al.} \cite{Albini} to represent quantum homodyne tomography in terms of a single
observable (as a normalized positive operator measure) we construct a generalized Markov kernel which transforms 
(the measurement outcome statistics of)
this observable into 
(the measurement outcome statistics of)
a covariant phase space observable. 
We also consider the inverse question. Finally,
we  add some remarks on the quantum theoretical justification
of the experimental implementations of these observables 
in terms of balanced homodyne  and 8-port detection techniques, respectively.
\end{abstract}
\maketitle

\section{Introduction}
There are two main approaches to continuous variable quantum tomography. The first one applies measurements of
rotated quadratures $\sfq_\theta$, $\theta\in[0,2\pi)$, the second one uses a measurement of the phase space observable 
${\mathsf G}_{\ket 0}$
generated by the vacuum state $\ket 0$.
In both cases the state $\rho$ of the measured system can actually be reconstructed from the corresponding measurement statistics,
either from the quadrature distributions
 $\rho^{\sfq_\theta}, \theta\in[0,2\pi)$, 
or from the phase space (Husimi) distribution $\rho^{{\mathsf G}_{\ket 0}}$ \cite{VR1989,Smithey1993,Dariano1994,LeonhardtII,Gianni2000,Paris_et_al}

A direct comparison of the two tomographic methods has suffered from the fact that in the first case 
one needs to measure, in principle, 
infinitely many different, mutually noncommuting  observables $\sfq_\theta, \theta\in[0,2\pi)$, whereas in the latter
approach a measurement of a single observable is sufficient.
For each state $\rho$
the random sampling of the  distributions $\rho^{\sfq_\theta}$, $\theta\in[0,2\pi)$, 
can, however,  easily be amalgamated into a probability bimeasure, 
which, when taken all together,
define a unique informationally complete positive operator measure $\Eh$, 
the \pom of the quantum homodyne tomography \cite{Albini}.

In this letter we construct a generalized Markov kernel which turns (the statistics of) the homodyne tomography observable $\Eh$ 
into (the statistics of) the phase space (Husimi) observable ${\mathsf G}_{\ket 0}$.
We also demonstrate that  there is no simple inverse construction. Finally, we briefly comment  on the 
theoretical background of the
physical implementation of the observables 
$\Eh$ and ${\mathsf G}_{\ket 0}$ via
balanced homodyne detection and 8-port homodyne detection tehniques, respectively.

\section{Notations and the two  POMs}
Let $h_n, n\in\N,$ denote the (normalized) Hermite functions in $\hi=L^2(\R)$, $N$
the corresponding number operator, and let
 $Q$ and $P$ stand for the usual position and momentum operators, respectively. 
Let $\sfq$ denote the spectral measure of $Q$
so that the quadrature observable $\sfq_\theta$ is  the spectral measure of the rotated  position operator $e^{i\theta N}Qe^{-i\theta N}$.
For any state $\rho$, positive trace one operator on $\hi$, we let $\rho^{\sfq_\theta}$ denote the density of the probability distribution
$X\mapsto \tr[\rho\,\sfq_\theta(X)]$, that is, $\tr[\rho\,\sfq_\theta(X)]=\int_X\rho^{\sfq_\theta}(x)\,dx.$

 For each state $\rho$,
the random sampling of the  distributions $\rho^{\sfq_\theta}$ 
can, indeed,  be amalgamated into a probability bimeasure 
 $\hB([0,2\pi))\times\br\ni(\Theta,X)\mapsto\int_\Theta \tr[\rho\,\sfq_\theta(X)]\frac{d\theta}{2\pi}\in[0,1]$,
which can uniquely be extended to a probability measure on $\hB([0,2\pi)\times\R)$ (we use the notation $\hB(\cdots)$ for the Borel sets).
As shown in \cite{Albini}, these probability measures define a unique
normalized positive operator measure, the \pom $\Eh$, with the property:\footnote{
Obviously, one could equally well define the \pom 
$\widetilde{\mathsf E}_{\rm ht}(\Theta\times X):=\frac{1}{\pi}\int_\Theta \Q_\theta(X)d\theta$  for  phase 
variable in $[0,\pi)$. Then
$
\Eh(\Theta\times X)=
\frac{1}{2}\widetilde{\mathsf E}_{\rm ht}\big(\Theta\cap[0,\pi)\times X\big)+\frac{1}{2}\widetilde{\mathsf E}_{\rm ht}\big([(\Theta\cap[\pi,2\pi))-\pi]\times (-X)\big)
$
for all $\Theta\in\bor{[0,2\pi)}$ and $X\in\bor\R$. To have a direct use of some of the results of  \cite{Albini} we choose to work with $\Eh$, which, however,
contains some redundancy in giving a double covering of $\R^2$.}
\begin{equation}
\Eh(\Theta\times X):=\frac{1}{2\pi}\int_\Theta \Q_\theta(X)d\theta
=\frac{1}{2\pi}\int_\Theta e^{i\theta N}\Q(X)e^{-i\theta N}d\theta,\ \ \Theta\in\bor{[0,2\pi)},\;X\in\bor\R.
\end{equation}
For any state $\rho$, we let $\rho^{\Eh}$ denote the
density of the probability measure $Z\mapsto \tr[\rho\Eh(Z)]$
with respect to $\frac 1{2\pi}d\theta dx$, so that $\rho^{\Eh}(\theta,x)=\rho^{\sfq_\theta}(x)$.
Since
the Radon transform of the Wigner function $\W_\rho$ of $\rho$ gives
the quadrature distributions, that is 
$(\Ra \W_\rho)(\theta,x)=\rho^{\sfq_\theta}(x)$, for almost all $\theta\in[0,2\pi), x\in\R$,
one thus has\footnote{The third equality is obtained whenever $\rho$ is such that $\W_\rho\in L^1(\R^2)$,
see, e.g. \cite{Albini}.}
\begin{eqnarray}
\tr[\rho\Eh(\Theta\times X)]&&=\frac 1{2\pi}\int_0^{2\pi}\int_{-\infty}^{\infty}\rho^{\Eh}(\theta,x)d\theta dx
=\frac 1{2\pi}\int_0^{2\pi}\int_{-\infty}^{\infty}\rho^{\sfq_\theta}(x)d\theta dx\\
&&=
\frac 1{2\pi}\int_0^{2\pi}\int_{-\infty}^{\infty}(\Ra\W_\rho)(\theta,x)d\theta dx.\nonumber
\end{eqnarray}

Let $W_{qp}=e^{\frac {pq}2}e^{-iqP}e^{ipQ}$ denote the unitary Weyl operators. 
As well known, see e.g.  \cite{HolevoII,WernerII,Cassinelli,NCQM},
any normalized \pom $\mathsf G$ which satisfies the
covariance condition with respect to the Weyl representation of the phase space translations,
has an operator density
$(q,p)\mapsto \frac 1{2\pi}W_{qp}KW_{qp}^*$
defined by a unique positive trace one operator $K$, that is, 
for any state $\rho$,
\begin{equation}
\tr[\rho {\mathsf G}(Z)]=\frac 1{2\pi}\int_Z\tr[\rho W_{qp}KW_{qp}^*]\,dqdq.
\end{equation}
In particular, if $K=\kb{h_0}{h_0}$, then the phase space observable ${\mathsf G}_K$
generated by $K$
 is denoted by ${\mathsf G}_{\ket 0}$, and the  density $\rho^{{\mathsf G}_{\ket 0}}$ of the
probability measure $Z\mapsto \tr[\rho {\mathsf G}_{\ket 0}(Z)]$ is just the Husimi $Q$-function of the state $\rho$, that is, $\rho^{{\mathsf G}_{\ket 0}}(q,p)= \frac 1{2\pi}\ip{W_{qp}h_0}{\rho W_{qp}h_0}$,
where $W_{qp}h_0=\ket z$ is the coherent state, with $z=\frac 1{\sqrt 2}(q+ip)$.

\section{Construction of the kernel}
We shall construct a generalized Markov kernel $M_{\ket 0}$ 
(defined on $\R^2\times\left([0,2\pi)\times\R\right)$)
such that for any state $\rho$, the density $\rho^{{\mathsf G}_{\ket 0}}$ is obtained from
the density $\rho^{\Eh}$ as follows:
\begin{equation}\label{m1}
\rho^{{\mathsf G}_{\ket 0}}(q,p)=
\frac 1{2\pi}\int_0^{2\pi}\int_\R
M_{\ket 0}^{q,p}(\theta,x)
\rho^{\Eh}(\theta,x)dx d\theta,
\end{equation}
or, equivalently,  for the   {\sf POM}s,
\begin{eqnarray*}
{\mathsf G}_{\ket 0}(Z)=
\int_0^{2\pi}\int_\R\underbrace{\left[\frac{1}{2\pi}
\iint _Z M_{\ket 0}^{q,p}(\theta,x)dqdp
\right]}_{=:\,{\cal M_{\ket 0}(Z;\theta,x)}
}
d\,{\Eh}(\theta,x)
=\int_0^{2\pi}\int_\R
{\cal M_{\ket 0}(Z;\theta,x)}
d\,{\Eh}(\theta,x).
\end{eqnarray*}
where $Z\subseteq\R^2$ is compact (so that we can change the order of integration).

In a forthcoming paper \cite{JP09} a general method is developed to construct such kernels
for  arbitrary phase space observables  ${\mathsf G}_{K}$.
This general method shows that the kernel can be obtained as a derivative of the Hilbert transform of the density of
the generating operator $K$.
 In the special case  of $K=\kb{h_0}{h_0}$ one may confirm this fact by a direct computation. We shall do so next.

Before doing that we note, however,  that essentially the same  computations have been carried out in a different
context of quantum-state sampling where the matrix elements of the state,
the density matrix,  are constructed form  the quadrature distributions in terms of a kernel function.  In that connection these functions
were called the sampling functions
or the pattern functions, see, e.g. \cite[Sect. 5]{LeonhardtII} and the references therein.

The Hilbert transform of the function $h_0^2$, 
$h_0(x) = \frac{1}{\sqrt[4]{\pi}}e^{-\frac 12 x^2}$,
can easily be computed and one observes that it is the Dawson integral
\begin{equation}
{\rm daw}(x)=
e^{-x^2}\int_0^x e^{t^2}dt.
\end{equation}
Following the method of  \cite{JP09} we now define the function
\begin{equation}\label{kernel}
M_{\ket 0}^{q,p}(\theta,x)
=2\frac{\partial}{\partial x}{\rm daw}(x-q\cos\theta-p\sin\theta),
\end{equation}
which is known \cite{KLP08} to be  a  bounded analytic function vanishing at infinity and having the values 
\begin{equation}\label{kerneli}
M_{\ket 0}^{q,p}(\theta,x)
=\sum_{k=0}^\infty \frac{(-1)^k k!}{2^k(2k)!}H_{2k}(x-q\cos\theta-p\sin\theta).
\end{equation}
It can easily be seen that $2\,d{\rm daw}(x)/dx$ is not a positive function
(see for instance the picture in page 114 of \cite{LeonhardtII})
 so that $\cal M_{\ket 0}(Z;\theta,x)$ is not a true conditional (or transition) probability; therefore we call $\cal M_{\ket 0}(Z;\theta,x)$ a {\it generalized} Markov kernel.

We go on to show that (for compact sets $Z$)
\begin{equation}\label{faasisuure}
{\mathsf G}_{\ket 0}(Z)=
\int_0^{2\pi}\int_\R
\left[\frac{1}{2\pi}
\int _Z M_{\ket 0}^{q,p}(\theta,x)dqdp
\right]
d{\Eh}(\theta,x).
\end{equation}
It suffices to show that the coherent state expectations of both sides of this equation are equal, that is, eq.\ (\ref{m1}) holds
for the coherent states $\rho=\kb zz$.
To verify this fact, 
we need some further notations.

For $(q,p)$ and $(u,v)$ in $\R^2$ we denote
$z=(q+ip)/\sqrt2$, $w=(u+iv)/\sqrt2$, and
$$
a:=q\cos\theta+p\sin\theta=\sqrt2 \, {\rm Re}(ze^{-i\theta}).
$$
If $\tilde w=we^{-i\theta}=(\tilde u+i\tilde v)/\sqrt2$ it follows that $\tilde u=u\cos\theta+v\sin\theta$ and
$$
y:=\tilde u-a=\sqrt2 \, {\rm Re}\big((w-z)e^{-i\theta}\big).
$$
Then
\begin{eqnarray*}
\<w|G_{|0\>}(Z)|w\>&=& \frac{1}{\pi}\int_Z\<w|z\>\<z|w\>d^2z=\frac{1}{2\pi}\int_Z e^{-|w-z|^2}dqdp,
\\
\<w|\Eh(\Theta\times X)|w\>&=&\frac{1}{2\pi}\int_\Theta\<we^{-i\theta}|\Q(X)|we^{-i\theta}\>d\theta,
\\
\<\tilde w |\Q(X)|\tilde w\>&=&\frac{1}{\sqrt{\pi}}\int_X e^{-(x-\tilde u)^2}dx.
\end{eqnarray*}
It thus suffices to show that
$$
e^{-|w-z|^2}=\frac{1}{\sqrt{\pi}}
\int_0^{2\pi}\int_\R M_{|0\>}^{q,p}(\theta,x) e^{-(x-\tilde u)^2}\textstyle{\frac{d\theta dx}{2\pi}}.
$$
Now
\begin{eqnarray*}
&&\frac{1}{\sqrt{\pi}}
\int_0^{2\pi}\int_\R M_{|0\>}^{q,p}(\theta,x) e^{-(x-\tilde u)^2}\textstyle{\frac{d\theta dx}{2\pi}} \\
&=& \frac{1}{\sqrt{\pi}}
\sum_{k=0}^\infty
\int_0^{2\pi}\frac{(-1)^k k!}{2^k(2k)!}\int _\R H_{2k}(x-a)e^{-(x-\tilde u)^2}\textstyle{\frac{ dxd\theta}{2\pi}}
\\&=&
\frac{1}{\sqrt{\pi}}
\sum_{k=0}^\infty
\int_0^{2\pi}\frac{(-1)^k k!}{2^k(2k)!}\underbrace{\int _\R H_{2k}(x)e^{-(x-y)^2}dx}_{=\,\sqrt{\pi}y^{2k}2^{2k}
\text{ (see \cite[7.374(6)]{taul}) }
}\textstyle{\frac{d\theta}{2\pi}}
\\&=&
\sum_{k=0}^\infty
\frac{(-1)^k k! 2^k}{(2k)!}\int_0^{2\pi}\underbrace{\left\{\frac{1}{2^k}\sum_{l=0}^{2k}{2k\choose l}
\left[(w-z)e^{-i\theta}\right]^l
\left[\overline{(w-z)}e^{i\theta}\right]^{2k-l}
\right\}}_{=\, y^{2k} \text{ since }y=\sqrt2 \, {\rm Re}\big((w-z)e^{-i\theta}\big)=2^{-1/2}
\big((w-z)e^{-i\theta}+\overline{(w-z)}e^{i\theta}\big)}
\textstyle{\frac{d\theta}{2\pi}}
\\&=&
\sum_{k=0}^\infty
\frac{(-1)^k k!}{(2k)!}
\sum_{l=0}^{2k}{2k\choose l}
(w-z)^l\overline{(w-z)}^{2k-l}\underbrace{\int_0^{2\pi}e^{2(k-l)}\textstyle{\frac{d\theta}{2\pi}}}_{=\,\delta_{l,k}}
\\&=&
\sum_{k=0}^\infty
\frac{(-1)^k k!}{(2k)!}\cdot\frac{(2k)!}{k!\,k!}|w-z|^{2k}=
\sum_{k=0}^\infty
\frac{\big(-|w-z|^2\big)^k}{k!}= e^{-|w-z|^2},
\end{eqnarray*}
showing that eq. (\ref{faasisuure}) is, indeed,  valid.

The construction of the Husimi \pom ${\mathsf G}_{\ket 0}$   from the homodyne detection observable $\Eh$ by the generalized Markov kernel (\ref{kerneli}),
as given in (\ref{faasisuure}), is based on a direct computation. Due to the informational completeness of the phase space observable ${\mathsf G}_{\ket 0}$ one would expect that
a similar reverse construction can also be given.  In the appendix we demonstrate, however, that no simple inverse construction can be given.

\section{Discussion}
The measurements of
the homodyne detection observable $\Eh$ and the phase space observable ${\mathsf G}_{\ket 0}$ constitute the basic measurements in the 
theory of (continous variable)
quantum tomography. Due to their informational completeness
their  measurement outcome statistics  $\rho^{\Eh}$ and $\rho^{{\mathsf G}_{\ket 0}}$ both separate states $\rho$. 
Together with the appropriate
inversion formulas the state $\rho$ can, indeed, be deduced both from  $\rho^{\Eh}$ as well as from $\rho^{{\mathsf G}_{\ket 0}}$. Therefore, the experimental
implementations of measurements of $\Eh$ and ${\mathsf G}_{\ket 0}$ as well as the inversion formulas 
$\rho^{\Eh}\mapsto\rho$ and $\rho^{{\mathsf G}_{\ket 0}}\mapsto\rho$ are most important, and they have widely been analysed in the literature. 
In spite of that we find it appropriate to add a few comments on these well-known methods and on the relevance of the kernel constructed above.

The balanced homodyne detection is well-known and much used techique to measure a quandarture observable $\sfq_{\theta}$, where $\theta$ is the phase
of the local oscillator, in coherent state $\ket z$, $z=|z|e^{i\theta}$, and $|z|>>1$. A rigorous quantum mechanical 
proof that the balanced homodyne detection observable tends, in the high amplitude limit $|z|\to\infty$, to the quadrature observable $\sfq_{\theta}$
has recently been worked out in \cite{JukkaVI}, but see also \cite{EV} and \cite{VG}.
The informational completeness of the quadrature observables $\sfq_{\theta}, \theta\in[0,\pi]$, is equally well-known; recent explicite proofs are given,
for instance, in \cite{Gianni2000,KLP08}. The statistical sampling of various $\sfq_{\theta}$ is, again, 
well-known, see e.g. \cite{LeonhardtII,Paris_et_al}, though the combination of them under a single observable $\Eh$ is a recent result \cite{Albini}. 
To our knowledge the only known measurement implementation of $\Eh$, however, is the high amplitude balanced homodyne detection together 
with a random change of the phase $\theta$.

Phase space observables ${\mathsf G}_K$, and, in particular, the Husimi \pom ${\mathsf G}_{\ket 0}$, are other important examples of informationally complete
observables. Indeed, ${\mathsf G}_K$ is informationally complete if and only if the generating operator $K$ is such that $\tr[W_{qp}K]\ne0$ for (almost all)
$(q,p)\in\R^2$. This results is due to \cite{Ali}, see \cite{KW} for some further technical elaborations. The Arthurs-Kelly model as well as the 8-port homodyne detection
scheme are, again, well-known measurement models for phase space observables ${\mathsf G}_K$, the latter one being also experimentally feasible. In fact, there
is a   recent rigorous proof that any phase space observable ${\mathsf G}_K$ can be obtained (as a high-amplitude limit of the local oscillator) of the 8-port homodyne
detection observable, with the generating operator $K$ defined by the state of the (single-mode) parametric field \cite{JukkaVII}.

The generalized Markov kernel  (\ref{kernel}) allows one to obtain the  Husimi distribution $\rho^{{\mathsf G}_{\ket 0}}$ of any state $\rho$ from the 
homodyne detection distribution $\rho^{\Eh}$ by the formula (\ref{m1}), that is, the phase space observable ${\mathsf G}_{\ket 0}$  from
the homodyne detection observable $\Eh$ 
by (\ref{faasisuure}).
This means, in particular, that any measurement  of ${\mathsf G}_{\ket 0}$ can, in principle, be reduced to
a measurement of $\Eh$.

\section*{Appendix: the reverse problem}
To solve the inverse problem,
we try to find a function $N_{|0\>}^{\theta,x}(u,v)$ such that
\begin{equation}\label{1}
\Eh(\Theta\times X)=
\iint_{\R^2}
\left[
\int_\Theta\int_X
N_{|0\>}^{\theta,x}(u,v)
d\theta d x
\right]\underbrace{d{\mathsf G}_{|0\>|}(w)}_{=\,|w\>\<w|dqdp/(2\pi)}
\end{equation}
where $w=(u+iv)/\sqrt2$, and where the order of integration can be changed (at least for compact sets $X\subset\R$).

As before, we denote
$z=(q+ip)/\sqrt2$, and $\tilde z=ze^{-i\theta}=(\tilde q+i\tilde p)/\sqrt2$ so that 
$\tilde q=q\cos\theta+p\sin\theta$, $\tilde p=-q\sin\theta+p\cos\theta$, 
$q=\tilde q\cos\theta-\tilde p\sin\theta$, $p=\tilde q\sin\theta+\tilde p\cos\theta$,
and $d\tilde qd\tilde p=dqdp$.

Operate on both sides of \eqref{1} by $\<-z|\cdots|z\>$. Now
$$
\<-z|\Eh(\Theta\times X)|z\>
=\frac{1}{2\pi}\int_\Theta \<-e^{-i\theta}z|\Q(X)|e^{-i\theta}z\>d\theta
$$
and
$$
 \<-e^{-i\theta}z|\Q_0(X)|e^{-i\theta}z\>= \<-\tilde z|\Q_0(X)|\tilde z\>
 =\frac{1}{\sqrt\pi}\int_X e^{i 2 \tilde p x-(x-\tilde q)^2/2-(x+\tilde q)^2/2}dx=\frac{1}{\sqrt\pi}\int_X e^{i 2 \tilde p x-x^2-\tilde q^2}dx.
$$
On the other hand,
$$
\<-z|{\mathsf G}_{|0\>}(Z)|z\>=\frac{1}{2\pi}\iint_Z\<-z|w\>\<w|z\>dudv
=\frac{1}{2\pi}\iint_Z e^{-|z|^2-|w|^2+2 i {\rm Im}(z\overline{w})}dudv.
$$
Thus, one must solve
$$
\frac{1}{\sqrt\pi}e^{i 2 \tilde p x-x^2-\tilde q^2} =
\iint_{\R^2} N_{|0\>}^{\theta,x}(u,v) e^{-|z|^2-|w|^2+2 i {\rm Im}(z\overline{w})}dudv
$$
or, equivalently, (note that $2 i {\rm Im}(z\overline{w})=i(pu-qv)$)
$$
\frac{1}{\sqrt\pi}e^{i 2 \tilde p x-x^2-\tilde q^2+|z|^2}=
\iint_{\R^2} N_{|0\>}^{\theta,x}(u,v) e^{-|w|^2}\cdot e^{i pu} e^{-i qv}dudv.
$$
If one makes assumptions that the right hand side of the above equation  (a function of $q$ and $p$) and the function
$
(u,v)\mapsto N_{|0\>}^{\theta,x}(u,v) e^{-(u^2+v^2)/2}
$
are integrable, one can take
take the double Fourier transform: one gets
$$
N_{|0\>}^{\theta,x}(u,v)e^{-(u^2+v^2)/2}=
\frac{1}{4\pi^2}\frac{1}{\sqrt\pi}\iint_{\R^2}e^{i 2 \tilde p x-x^2-\tilde q^2+|z|^2}\cdot e^{-i pu} e^{i qv}dqdp.
$$
If the last integral would exist, then one could change integration variables to get
\begin{eqnarray*}
\iint_{\R^2}e^{i 2 \tilde p x-x^2-\tilde q^2+|z|^2}\cdot e^{-i pu} e^{i qv}dqdp
&=&
\iint_{\R^2}e^{i 2 \tilde p x-x^2-\tilde q^2+|\tilde z|^2-i(\tilde q\sin\theta+\tilde p\cos\theta)u+
i(\tilde q\cos\theta-\tilde p\sin\theta)v}d\tilde qd\tilde p
\end{eqnarray*}
where the argument of the exponential function of the integrand is
\begin{eqnarray*}
&&i 2 \tilde p x-x^2-\tilde q^2+|\tilde z|^2-i(\tilde q\sin\theta+\tilde p\cos\theta)u+i(\tilde q\cos\theta-\tilde p\sin\theta)v \\
&&=-\frac{1}{2}\tilde q^2+\frac{1}{2}\tilde p^2+i\tilde q(-u\sin\theta+v\cos\theta)+i\tilde p(2x-u\cos\theta-v\sin\theta)-x^2.
\end{eqnarray*}
Certainly the above integral does not exist even as a tempered distribution due to the $\frac12\tilde p^2$-term;
for example, if $u=v=x=0$, then one would have
$$
N_{|0\>}^{\theta,0}(0,0)=\frac{1}{4\pi^2}\frac{1}{\sqrt\pi}
\int_\R e^{\frac{1}{2}\tilde p^2}d\tilde p
\int_\R e^{-\frac{1}{2}\tilde q^2}d\tilde q
$$
which clearly does not exist.
To conlude, there is no function $N_{|0\>}^{\theta,x}(u,v)$ with the above properties which would allow one to write
$
\Eh(\Theta\times X)=
\iint_{\R^2}
\left[
\int_\Theta\int_X
N_{|0\>}^{\theta,x}(u,v)
d\theta d x
\right]d{\mathsf G}_{|0\>|}(w)$.


\begin{thebibliography}{99}
\bibitem{Albini} P. Albini, E. De Vito, A. Toigo, Quantum homodyne tomography as an informationally complete positive operator
valued measure, arXiv:0807.3437 (July 2008). 
\bibitem{Ali} S.T. Ali, E. Prugove\v cki, Classical and quantum statistical mechanics in a 
common Liouville space,
{\em Physica} {\bf 89A} (1977) 501-521.
\bibitem{Gianni2000} G. Cassinelli, G.M. D'Ariano, E. De Vito, A. Levrero, {\em Group theoretical quantum tomography},
{\em J. Math. Phys.} {\bf 41} (2000) 7940-7951.
\bibitem{Cassinelli} G. Cassinelli, E. De Vito, A. Toigo, Positive operator valued measures covariant with respect to an irreducible representation,
{\em J. Math. Phys.} {\bf 44} (2003) 4768-4775.
\bibitem{Dariano1994} G. M. D'Ariano, C. Macchiavello, M. G. A. Paris, Detection of the density matrix through optical homodyne
 tomography without filtered back projection, {\em Phys. Rev. A} {\bf 50} (1994) 4298.
 \bibitem{taul} I. S. Gradshteyn, I. M. Ryzhnik, {\em Table of Integrals, Series, and Products},
Corrected and Enlarged Edition, Academic Press, Inc., Orlando, 1980.
\bibitem{HolevoII} A. S. Holevo, Covariant measurements and uncertainty relations, {\em Rep. Math. Phys.} {\bf 16} (1979) 385-400.
\bibitem{JukkaVI} J. Kiukas, P. Lahti, On the moment limit of quantum observables, with an application to the balanced homodyne detection, 
{\em J. Mod. Optics} {\bf 55}  (2008) 1175-1198.
\bibitem{JukkaVII} J. Kiukas, P. Lahti, A note on the measurement of phase space observables with an eight-port homodyne detector, 
{\em J. Mod. Optics} (2007)
\bibitem{KLP08} J. Kiukas, P. Lahti, J.-P. Pellonp\"a\"a,
A proof for the informational completeness of the rotated quadratures, {\em J. Phys. A: Math. Theor.} {\bf 41} (2008) 175260 (11p).
\bibitem{NCQM} J. Kiukas, P. Lahti, K. Ylinen, Normal covariant quantization maps, {\em J. Math. Anal. Appl.} {\bf 319} 783-801 (2006). 
\bibitem{KW} J. Kiukas, R. Werner, private communication.
\bibitem{JP09} J.-P. Pellonp\"a\"a, to be published.
\bibitem{LeonhardtI} U. Leonhardt, H. Paul, Phase measurement and Q function, {\em Phys. Rev. A} {\bf 47} (1993) R2460-R2463.
\bibitem{LeonhardtII} U. Leonhardt, {\em Measuring the Quantum State of Light}, Cambridge University Press, Cambridge, 1997.
\bibitem{Paris_et_al} M. Paris, J. \v Reh\' a\v cek (Eds), {\em Quantum State Estimation}, Lect. Notes Phys. {\bf 649}, Springer-Verlag, Berlin, 2004.
\bibitem{Smithey1993} D.T. Smithey, M. Beck, M.G.Raymer, A. Faridina, Measurement of the Wigner distribution and the density 
matrix of a light mode using optical homodyne tomography: application to squeezed states and the vacuum,
{\em Phys. Rev. Lett.} {\bf 70} (1993) 1244-1247.
\bibitem{EV} E. Vogel, {\em Operationale Untersuchung von quantenoptischen Me\ss prozessen}, Shaker Verlag, K\"oln 1996.
\bibitem{VR1989} K. Vogel, H. Risken, Determination of quasiprobability distributions in terms of probability distributions for
the rotated quadrature phase, {\em Phys. Rev. A} {\bf 40} (1989) 2847-2849.
\bibitem{VG}
W. Vogel, J, Grabow,
Statistics of difference events in homodyne detection
{\em Phys. Rev. A}  {\bf 47} (1993) 4227 - 4235.
\bibitem{WernerII} R. Werner, Quantum harmonic analysis on phase space, {\em J. Math. Phys.} {\bf 25} (1984) 1404-1411 .





\end{thebibliography}
\end{document}